\documentclass[12pt,preprint]{aastex}

\shorttitle{Low Frequency 3C~391}
\shortauthors{Brogan et al.}

\begin{document}
\newcommand\HII{H\,{\sc ii}}
\newcommand\HI{H\,{\sc i}}
\newcommand\OI{[O\,{\sc i}] 63 $\mu$m}
\newcommand\CII{[C\,{\sc ii}] 158 $\mu$m}
\newcommand\CI{[C\,{\sc i}] 370 $\mu$m}        
\newcommand\SiII{[Si\,{\sc ii}] 35 $\mu$m}
\newcommand\Hi{H110$\alpha$}
\newcommand\He{He110$\alpha$}
\newcommand\Ca{C110$\alpha$}
\newcommand\kms{km~s$^{-1}$}
\newcommand\cmt{cm$^{-2}$}
\newcommand\cc{cm$^{-3}$}
\newcommand\mum{$\mu$m}
\newcommand\muG{$\mu$G}
\newcommand\mjb{mJy~beam$^{-1}$}
\newcommand\jb{Jy~beam$^{-1}$}
\newcommand\dv{$\Delta v_{FWHM}$}
\newcommand\va{$v_A$}
\newcommand\Np{$N_p$}
\newcommand\np{$n_p$}
\newcommand\pp{^{\prime\prime}}
\newcommand\km{km~s$^{-1}$}
\newcommand\h{^{\rm h}}
\newcommand\m{^{\rm m}}
\newcommand\s{^{\rm s}}
\newcommand{\et}{\it et al.\rm}

\title{Spatially Resolved Low Frequency VLA observations of the \\
Supernova Remnant 3C 391}

\author{C. L. Brogan\altaffilmark{1,2}, 
T. J. Lazio\altaffilmark{3},  N. E. Kassim\altaffilmark{3}, and
K.  K. Dyer\altaffilmark{3,4}}

\altaffiltext{1}{University of Hawaii, Institute for Astronomy, 
640 North A'ohoku Place, Hilo, HI 96720; cbrogan@ifa.hawaii.edu}

\altaffiltext{2}{James Clerk Maxwell Telescope Postdoctoral Fellow}

\altaffiltext{3}{Naval Research Laboratory, Remote Sensing Division, 
Code 7213, 4555 Overlook Avenue SW, Washington, DC 20375-5351; 
joseph.lazio@nrl.navy.mil; namir.kassim@nrl.navy.mil; kdyer@nrao.edu}

\altaffiltext{4}{NRC Postdoctoral Fellow}

\begin{abstract}

We present VLA images of the supernova remnant (SNR) 3C~391 at 74,
330, and 1465 MHz.  This remnant has been known for some time to
exhibit a turnover in its integrated radio continuum spectrum at
frequencies $<$ 100 MHz, indicative of free-free absorption from
thermal ionized gas along the line of sight.  For the first time, our
data reveal the spatially resolved morphology of the low frequency
free-free absorption with a resolution of $\sim 70\arcsec$.  Contrary
to the expectation that such absorption arises from unrelated low
density \HII\/ regions (or their envelopes) along the line of sight,
these data suggest that in this case the absorbing medium is directly
linked to the SNR itself.  3C~391 has been shown in a number of recent
papers to be interacting with a molecular cloud.  Indeed, it exhibits
a number of signposts of SNR/molecular cloud shocks including OH (1720
MHz) masers and broad molecular emission lines.  Comparison of the
regions of strongest 74~MHz absorption with existing X-ray, IR, and
molecular data suggests that the free-free absorption originates from
the SNR/molecular cloud shock boundaries due to ionized gas created
from the passage of a J-type shock with a speed of $\sim 100$ \kms\/.
This makes only the second SNR for which such (extrinsic) spatially
resolved absorption has been measured, and the only one for which the
absorption is thought to arise from a SNR/molecular cloud interface
region.

\end{abstract}


\section{INTRODUCTION}

Supernovae have a profound effect on the morphology, kinematics, and
ionization balance of galaxies.  The impact of supernova shocks on
surrounding molecular clouds may also trigger new generations of star
formation. The idea that cosmic-rays (up to $\sim 10^{14}$ eV) are
produced by shock acceleration in supernova remnants (SNRs) has
recently gained wide acceptance \citep[e.g.][]{Jones1998}.  Although
the details of the SNR cosmic-ray acceleration process remain
uncertain, many of the current models predict spatial variations in
the spectral indices of SNRs at the sites of cosmic-ray acceleration
\citep[c.f.][]{Reynolds1992}.  The spectral index variations predicted
by many of these models are most easily discerned at low radio
frequencies. There is also evidence that some of the low latitude
unidentified EGRET $\gamma$-ray sources may be associated with sites
of SNR/molecular cloud interactions, implying that these sources are
particularly efficient at producing high energy cosmic-rays \citep[see
for example][]{Torres2003}. Thus, sites of SNR/molecular cloud
interaction represent a promising place to look for signatures of
shock acceleration. However, unambiguous evidence for SNR/molecular
cloud interactions is difficult to find due to Galactic velocity
confusion in the low-$J$ molecular lines generally used for such
searches, so that only a handful of such sources are currently known
\citep[c.f.][]{Reach2002}.

A further complication is the determination of whether observed
spectral index variations are intrinsic to the synchrotron emission, or
are affected by other processes.  Using integrated flux measurements,
\citet{Kassim1989b} found that $\sim 2/3$ of Galactic SNRs undergo a
spectral turnover for frequencies $\lesssim 100$ MHz.  These low
frequency turnovers have been attributed to free-free absorption in
low density $n_e\sim 1 - 10$ \cc\/ intermediate temperature ($T\sim
5000$ K) ionized thermal gas along the line of sight to the SNR. The
ubiquitous detection of low frequency radio recombination lines (RRLs)
amplified by stimulated emission, towards almost every direction in
the inner Galaxy
\citep[e.g.][]{Roshi2001,Anantharamaiah1986,Anantharamaiah1985}
provides further evidence for ionized gas with the requisite densities
and temperatures.  The nature of the ionized gas remains unclear, with
the extended envelopes of normal \HII\/ regions being one candidate
\citep[also see \S4.3,][and references therein]{Brogan2004}.
Spatially resolved low frequency observations of SNRs with absorption
will be crucial for understanding the origin of the absorbing
medium. Also, while such absorption is interesting in its own right
for understanding the composition of the ISM, it hinders efforts to
look for unambiguous spectral index variations predicted by cosmic-ray
acceleration models.  Therefore, quantifying the level and origin of
free-free absorption toward SNRs is also critical to studies of the
cosmic-ray shock acceleration process.

The first spatially resolved observations of extrinsic low frequency
absorption toward an SNR were made by \citet{Lacey2001} toward W49B
using the then recently completed 74~MHz system at the Very Large
Array (VLA). These authors found that most of the low frequency
absorption implied by previous low resolution integrated flux
measurements is confined to the western half of the remnant, and is
spatially coincident with a neutral \HI\/ column density peak at a
kinematic distance well in front of the SNR.  Evidence for the
requisite ionized gas comes from radio recombination line (RRL)
detections at the same velocity as the neutral \HI\/ gas.  Thus, the
synchrotron emission from W49B is absorbed by an unrelated partially
ionized cloud along the line of sight.

Based on the interesting W49B result, and recent improvements in low
frequency widefield imaging techniques we have begun a
low frequency survey of the inner Galactic plane with the
VLA\footnote{The National Radio Astronomy Observatory is a facility of
the National Science Foundation operated under cooperative agreement
by Associated Universities Inc.} at 74 and 330~MHz.  With this study
we hope to spatially resolve and characterize the low frequency
absorption toward a large sample of SNRs from $\ell= -10\arcdeg$ to
$50\arcdeg$ \citep[see][and Brogan et al.  2004, in prep]{Brogan2004}.
This paper presents results from this survey at 74~MHz as well as
archival VLA data at 330 and 1465 MHz for the SNR 3C 391
(G31.9+0.0). The organization of the paper is as follows: \S1.2
describes previous observations of 3C~391, \S2 briefly describes the
data reduction, \S3 describes our results, \S4 presents comparisons of
our data with previous observations toward 3C~391 and discusses the
implications, and our findings are summarized in \S5.

\subsection{Previous observations of 3C~391}

From integrated radio flux measurements, 3C~391 has been known from
the 1970s to suffer from low frequency absorption \citep[see for
example][]{Dulk1975}. In addition, RRLs (which cannot originate from
the synchrotron emission of the SNR) have been detected toward 3C~391
at velocities of $\sim 100$ \kms\/ \citep{Pankonin1976, Cesarsky1973}.
\citet{Reynolds1993} and \citet{Moffett1994} report the results of the
first high resolution radio continuum observations toward 3C~391 at
330, 1468, and 4848 MHz (earlier, lower resolution radio frequency
studies are also reviewed in these references). At high resolution,
3C~391 displays a rim brightened morphology to the NW with the
synchrotron emission gradually fading to the SE, indicative of a
``breakout'' into lower density gas in this region.
\citet{Moffett1994} hoped to find spectral signatures of shock
acceleration in the resulting high resolution spectral index maps of
3C~391, but could find no convincing evidence for such variations
above their detection limit of $\Delta\alpha=0.1$ ($S\propto \nu^{\alpha}$).  

It has recently been discovered that 3C~391 is interacting with a
nearby molecular cloud which also has a velocity of $\sim 100$ \kms\/
\citep{Wilner1998}. Further proof of an interaction comes from the
presence of OH (1720 MHz) masers which trace SNR/molecular cloud
shocks \citep[when the mainline masers are absent;][] {Frail1996}.
More detailed observations of the associated molecular gas have been
carried out by \citet{Reach1999}. These authors find evidence for
broad CO linewidths indicative of a shock at the location of one of
the OH masers.  The distance to the SNR constrained by the velocity of
maximum \HI\/ absorption (which is equal to that of the tangent
point), CO emission line velocities, and the velocities of the OH
masers ($\sim 105$ \kms\/) is $\sim 7.5$ kpc
\citep{Radhakrishnan1972,Wilner1998,Frail1996}.  At a distance of 7.5
kpc, $1\arcmin$ is equivalent to 2.2~pc. Extensive near and mid
infrared observations of 3C~391 have been carried out by
\citet{Reach2002} and \citet{Reach2000} which have yielded further
detailed information on the nature of the shock interaction. The
results of these studies are discussed further in \S 4.2.

The X-ray properties of 3C~391 have been studied by \citet{Chen2004},
\citet{Chen2001}, \citet{Rho1996}, and \citet{Wang1984} using {\em
Chandra}, ASCA, ROSAT, and {\em Einstein} respectively. Overall, the
X-ray morphology of 3C~391 places it in the ``mixed morphology'' or
``thermal composite'' class \citep{Rho1998}. That is, the X-ray
emission is brightest interior to the radio synchrotron shell, with
weak or no emission coincident with the radio shell. In all of the
X-ray studies described above, the centroid of soft X-ray emission is
offset to the SE of the harder X-ray emission. This has been
interpreted as an absorption effect due to the extra column density
provided by the molecular cloud to the NW \citep{Chen2004,Chen2001}.
\citet{Chen2004} find that the column density difference implied by the
soft x-ray absorption is $\sim 6\times 10^{21}$ \cmt\/. These authors
also suggest that if the column density difference is due to the
interacting molecular cloud, the implied {\em average} molecular cloud
density is $\sim 20$ \cc\/ where they have assumed a cloud depth of
$\sim 12$ pc ($5\arcmin$ at a distance of 8 kpc, comparable to the
angular extent of 3C~391). Alternative estimates for the cloud density 
are presented in \S4.3.

\section{OBSERVATIONS}

We observed the supernova remnant 3C~391 using the Very Large Array
(VLA) at 74 MHz using the A, B, \& C configurations. In addition, we
obtained 330 C configuration and 1465 MHz D configuration data from
the VLA archive. This combination of configurations provides
comparable $u-v$ coverage at all three frequencies -- a prerequisite
for believable spectral index calculations. The observing parameters
for each frequency are presented in Table 1.  The 74 and 330~MHz data
were reduced using the wide-field imaging and calibration techniques
described in \citet[][, and references therein]{Brogan2004}.  The 1465
MHz data were reduced in the usual manner. All data reduction utilized
the AIPS software package.

The final resolution of the 74 MHz image after combining the four
datasets described in Table 1 is $70\arcsec$. This resolution is about
2.8 times larger than what is technically feasible from the longest A
configuration baseline (30 km). This is due in part to the fact that
the A-configuration 74~MHz data were taken under poor ionospheric
conditions which caused the longest baselines to have incoherent
phases. To mitigate this effect the 74~MHz data were imaged with a
$u-v$ taper of 5.5 k$\lambda$ (this corresponds to a baseline length
of 22 km). In addition, the combination of A+B+C configuration data
pushes the final resolution to one that is intermediate (B) between
the configuration extremes (A and C). The 330 and 1465 MHz images were
convolved to match the $70\arcsec$ of the 74 MHz images. This
convolution was minimal for both the C configuration 330 MHz and D
configuration 1465 MHz data (each with an uncovnvolved resolution of
$\sim 60\arcsec$ resolution).
 
Since 3C~391 is located $2.2\arcdeg$ from the phase center of the 74
MHz data, primary beam correction was applied to this image (the
radius of the FWHM primary beam of the VLA at 74~MHz is $\sim
5.8\arcdeg$).  No primary beam correction was necessary at 330 and
1465 MHz since 3C~391 subtends a small fraction of the primary beam at
these frequencies and 3C~391 was located at the phase center of these
observations.  The rms noise levels achieved in each image are: 75,
15, and 3 \mjb\/ at 74, 330, and 1465 MHz, respectively. Note that
these rms noise levels are significantly higher than theoretical, due
to imperfect $u-v$ coverage (especially for the archival data) and
Galactic confusion.

\section{RESULTS}

Figure 1a shows our 330~MHz $70\arcsec$ resolution continuum image of
3C~391, while Figure 1b shows a 330/1465 MHz spectral index map of
3C~391.  Even at this fairly coarse resolution it is apparent that
3C~391 has a rim-brightened morphology on its western side and has a
breakout morphology to the SE indicating that the remnant is unbounded
in this region \citep[see][for higher resolution
images]{Moffett1994}. Figure 1c shows the notably different morphology of 
the 74~MHz emission, while Figure 1d shows the  74/330~MHz spectral
index. For both spectral index maps, the images
were masked at the $4\sigma$ level before the spectral index was
calculated (we use the definition $S\propto \nu^{\alpha}$). Neglecting
edge effects (the spectral index maps become more uncertain with
decreasing continuum brightness), the 330/1465 MHz spectral index map
is quite uniform with an average spectral index of $-0.45$.  The 74/330
MHz spectral index map is dramatically different from the 330/1465 MHz
map. Indeed the 74/330~MHz spectral indices are inverted to positive
values in a ring like morphology around the SNR. Such spectral
behavior is contrary to that expected of nonthermal emission and is
indicative of free-free absorption at 74~MHz, the morphology of which
is revealed here for the first time.

Integrated flux density measurements were made for the three
frequencies in our study for 3C~391.  Since the images have different
noise levels, and 3C~391 has a ``breakout'' morphology to the SE that
gradually fades into the noise it was difficult to determine where the
boundary of the SNR should be drawn.  Therefore, in order to get
comparable flux density estimates, the $4\sigma$ (60 \mjb\/) contour
of the 330~MHz image (see Figure 1a) was used to define the boundary
for all of the integrated flux measurements.  The integrated flux
densities are $20.2\pm 0.1$, $38.9\pm 0.3$, and $28.1\pm 1.8$ Jy at
1465, 330, and 74~MHz, respectively.  The errors on these flux density
measurements were estimated by ${\rm
(\#~independent~beams)^{1/2}}\times 3\sigma$ at each frequency.

A plot of the integrated continuum spectrum of 3C~391 is shown in Figure 2 
including data from the current work and integrated flux densities
from the literature for cases in which an error estimate is available
and the error is less than 20\%. As described in \S 1, it has been
known for some time that 3C~391 has a low radio frequency turnover,
and this effect is quite apparent in Figure 2 for $\nu< 100$ MHz.  In
order to determine the properties of the turnover, we have made a
weighted least squares fit of the integrated flux densities shown in
Fig. 2 to the equation
\begin{equation}
S_{\nu}= S_{330}\left(\frac{\nu} {330~{\rm MHz}}\right)^{\alpha}
~{\exp}\left[-\tau_{330} \left(\frac{\nu} {330~{\rm
MHz}}\right)^{-2.1}\right],
\end{equation}
where $S_{330}$ and $\tau_{330}$ are the flux density and optical
depth at a fiducial frequency of 330~MHz, respectively, and $\alpha$
is the integrated spectral index \citep{Dulk1975, Kassim1989a}.  This
equation assumes a standard nonthermal constant power law spectrum and
allows for a thermal absorption turnover at lower frequencies
\citep[see e.g.][]{Dulk1975, Kassim1989a}.  The free-free continuum
optical depth at other frequencies can be estimated from
$\tau_{\nu}=\tau_{330} \left(\frac{\nu} {330~{\rm
MHz}}\right)^{-2.1}$. The best fit parameters using the data shown in
Figure 2 are $S_{330}=41.0$ Jy, $\alpha=-0.49\pm 0.1$ and an average
optical depth at 74~MHz of $\tau_{74}=1.1$. The average optical depth
at 330~MHz is 0.08, indicating that the emission at this frequency is
little affected by free-free absorption, although the average 330/1465
MHz spectral index ($-0.45$) is slightly shallower than that
calculated from the full spectrum ($-0.49$) due to this effect.

The derived value of the integrated $\tau_{74}=1.1$ is about a factor
of two larger than that predicted by \citet{Kassim1989b}. This
discrepancy is due in large part to the fact that the
\citet{Kassim1989b} fit includes a $3\sigma$ 30.9 MHz {\em
non-detection} upper limit, whereas we have not included this data
point in our fit (although this upper limit does appear on Fig. 2 for
reference).  In addition, we also use a weighted least squares fit
which places greater emphasis on the more precise flux measurements,
including the three new flux measurements from this study. From Figure
2 it is clear that the flux density of 3C~391 at 30.9 MHz is likely to
be significantly lower than the $3\sigma$ 30.9 MHz detection limit of
the \citet{Kassim1989b} data.  Thus, we are confident that this
increase in the $\tau_{74}$ estimate is real, but an even lower
frequency {\em detection} (i.e. $< 74$ MHz) would better constrain the
fit.

\section{DISCUSSION}

\subsection{Comparison to X-ray morphology}

Figure 3a shows hard X-ray emission contours integrated from 2.6 - 10
keV from ASCA \citep{Chen2001} superposed on the 74/330~MHz spectral
index map. The Hard X-rays are little affected by absorption and
indicate the full extent of the SNR. Soft X-ray contours integrated
from 0.5-2.6 keV from ASCA \citep{Chen2001} superposed on the 74/330
MHz spectral index map are presented in Figure 3b. In contrast to the
hard X-rays, the soft X-rays are dramatically absorbed toward the
western side of the SNR. This is indicative of an enhanced column
density of hydrogen gas (both molecular and neutral) toward the NW and
western sides of the SNR compared to the SE as discussed in \S1.1. The
region of higher column density (indicated by regions with hard X-rays
but no soft X-rays) is well correlated with that of the deepest 74~MHz
absorption.

\subsection{Comparison to molecular and atomic line emission}

Figure 3c shows CO (2$-$1) integrated emission contours from 91 to 110
km/s tracing the molecular cloud that is interacting with 3C 391
\citep{Reach1999} superposed on the 74/330~MHz spectral index map. The
OH (1720 MHz) masers discovered by \citet{Frail1996} are also
indicated for reference. Much of the CO ($2-1$) emission shown in
Figure 3c has narrow line widths, indicating that it has not yet been
affected by the shock. Only toward the CO emission peak (near the
southern OH maser source) is there obvious evidence for shocked CO gas
\citep{Reach1999}.  Figure 3d shows mid infrared emission contours
from ISOCAM at 12-18 \mum\/ from \citet{Reach2002} superposed on the
74/330~MHz spectral index map.

There is impressive agreement between the regions of strongest 74~MHz
absorption and the 12-18 \mum\/ IR emission shown in Fig. 3d.  Such a
correlation may seem somewhat unexpected since dust cannot be
responsible for the free-free absorption. Indeed, \citet{Reach2002}
found that the bulk of the infrared emission in the 12-18 \mum\/ range
toward 3C 391 arises from fine structure atomic [NeII] and [NeIII]
infrared lines. These authors also found that the only regions that do
have significant dust emission in the 12-18 \mum\/ range lie outside
of the radio shell (i.e. to the SW and NE; Fig. 3d). In addition to
the [NeII] and [NeIII] lines in the 12-18 \mum\/ range,
\citet{Reach2002} also find a similar distribution for [FeII] at 1.64
\mum\/ (albeit over a smaller field of view encompassing only the
south and western parts of the remnant).  Ionized infrared fine
structure lines have also been shown to account for much of the 12-25
\mum\/ emission observed toward the northeast rim of IC~443; another
SNR known to be interacting with a molecular cloud \citep{Oliva1999b}.
 
Infrared fine structure lines can be easily formed in the cooling
post-shock gas after the passage of a dissociating J-type shock with a
speed $\sim 100$ \kms\/ \citep[see for example][]{Hollenbach1989}.  
Indeed, \citet{Reach2002} suggest that since
shocked molecular gas as traced by wide CO line widths and excited H$_2$
emission is concentrated near the CO peak (Fig. 3c), while the IR line
emission is much more widespread (the morphology agrees well with the
radio continuum), that much of the shock interaction between the SNR
and molecular must be a dissociative J-type shock. This
interpretation is supported by the lack of mid-IR continuum or
hydrocarbon line emission interior to the remnant, since much of the
dust would be destroyed in such a shock (the high abundance of gas
phase [Fe] also implies that most of the dust has been destroyed).
Thus, the 12-18 \mum\/ emission predominately traces the SNR/molecular
cloud shock boundaries.

\subsection{Properties of Free-Free Absorbing Gas}

Using Equation 1, and the integrated spectral index found in 
\S3.1 of $-0.49$, we derive the following equation for the 74~MHz 
optical depth as a function of position
\begin{equation}
\tau_{74}(\alpha,\delta)= -{\rm ln}\left[0.481\left(\frac{S_{74}(\alpha,\delta)} {S_{330}(\alpha,\delta)}\right)\right],
\end{equation}
where $S_{74}(\alpha,\delta)$ and $S_{330}(\alpha,\delta)$ are the
images shown in Figures 1a and 1c. The resulting 74~MHz optical depth
map is shown in Figure 4. The morphology of the $\tau_{74}$ map mimics
that of the 74/330~MHz spectral index maps shown in Figures 1 and 3.
If the electron temperature is known, the $\tau_{74}$ map 
shown in Figure 4 can be converted a map of emission measure ($EM$) using
\begin{equation}
EM=6.086\times 10^{-6}~a(T_e,\nu)^{-1}~\nu^{2.1}~\tau(\nu)~T_e^{1.35} ~~ 
{\rm cm}^{-6} pc,
\end{equation}
where $EM$ is the emission measure, $T_e$ is the electron temperature
(K), $\nu$ is the frequency in MHz, and $a(T_e,\nu)$ is the Gaunt
factor and is $\sim 1$ at the temperatures and densities discussed
here \citep{Dulk1975}. 

In the remainder of this section we use previous observational data
toward 3C~391 (see \S4.1 and 4.2) to constrain the properties of the
free-free absorbing gas. However, it is important to note that the
electron temperature and density (both electron $n_e$ and total $n$)
will change significantly as a function of distance behind the shock
front \citep[see for example][]{Hollenbach1989}. Moreover, since we
are observing at least some of the shock front ``face-on'' (see Fig 3)
we are sampling all of these regions along the line of sight
simultaneously in a way that cannot be disentangled. Thus, it is
impossible to uniquely assign any one value to either $T_e$ or
$n_e$. For the remainder of this section we deal only with average or
typical values, in order to give a sense of the likely parameter
space.

In order to account for the wide range of near infrared, mid-infrared,
and molecular lines observed toward 3C~391, \citet{Reach2000} propose
a three component model for the pre-shock gas: atomic (A; $n_o\sim 1$
\cc\/), molecular (M; $n_o\sim 100$ \cc\/), and dense clump (C;
$n_o\sim 10^4$ \cc\/). {\em After passage of the dissociating J-shock},
the primary tracers of each of these components are A: NeIII and OIII
($n\sim 10$ \cc\/), M: OI, FeII, SiII ($n\sim 10^3$ \cc\/), and C: H$_2$,
CS(7-6), and OH (1720 MHz) masers \citep[$n\sim 10^5$
\cc\/;][]{Reach2000,Reach2002}.  Based on the observed properties of the
un-shocked molecular cloud outside of the boundaries of the SNR
\citep{Reach1999}, it seems likely that the majority of the pre-shock
gas was composed of component ``M'' gas.  As noted by
\citet{Reach2000}, the gas will not of course be segregated into these
distinct categories -- instead there will be a continuum of these
conditions.  However, using a number of diagnostics, these authors
find that this simple three component model does a good job of
reproducing the majority of the observed spectral lines. 
 
Given the excellent morphological agreement between the fine structure
ionized atomic line emission and the 74 MHz free-free absorption
described in \S4.2 (Figure 3d), it seems reasonable to assume that the
two tracers coexist. That is that the electrons responsible for the
free-free absorption are those created from the ionization of the fine
structure line emitting atoms like FeII.  By comparing a number of
different near and mid-infrared [FeII] lines, \citet{Reach2002} find
that the line intensities of the iron lines are best fit by a
temperature $\lesssim 3,000$ K.  Using this temperature and Equation
3, the average 74~MHz optical depth of 1.1, and $T_e=1,000~-~3,000$ K
we find that the average $EM= (0.6 - 2.5)\times 10^3$ cm$^{-6}$
pc. However, it is not clear that the ions and electrons would be
thermalized as suggested here, especially since the ionization
fraction is unknown. In lieu of a better estimate we will assume that
$T_e=1,000~-~3,000$ K, but the uncertainty of this assumption is
inherent in the remaining calculations that use the average $EM$.

Likewise, it seems reasonable to assign the post-shock density that is
appropriate for the FeII lines of $n\sim 10^3$ \cc\/ to the free-free
absorption region (i.e. according to the three component model
described above). However, the post-shock electron density ($n_e$) is
the key parameter for the free-free absorption and is unknown.  In the
\citet{Reach2002} study, little evidence was found for {\em shocked}
molecular gas (component D, due to dense clumps) with the exception of
the region near the southern OH maser (see Fig. 3c). Additionally,
little mid-infrared continuum emission was observed in the ISO data,
implying wide scale destruction of the dust by the passage of the
shock. This conclusion is also supported by the presence of strong
gas-phase Fe and Si lines since these species are normally depleted
onto grains. From these clues, these authors suggest that much of
molecular gas must also have been destroyed by the shock, although
there may still be a significant amount of neutral atomic gas present.
Indeed, the presence of widespread OI suggests that not everything has
been even singly ionized, thus we view $n\sim 10^3$ \cc\/ as a strict
upper limit to the electron density \citep[also see][]{Oliva1999a}.
It is also possible that some of the free-free absorption arises from
the ``A'' zone (traced by OIII and NeIII), but since we expect
$n_e\sim n\sim 10$ \cc\/ in this region \citep[see above
and][]{Reach2000}, our $n_e\la n\sim 10^3$ \cc\/ upper limit is not
violated.

Using $EM=\int n_e^2~d\ell$, $n_e\la n\sim 10^3$ \cc\/ as an upper
limit for the electron density, and the emission measure derived
above, we find that the thickness of the ionized gas layer responsible
for the free-free absorption is $\gtrsim 0.0006-0.0025$ pc. It is
notable that some of the free-free absorption could also arise from
the ``A'' zone (traced by OIII and NeIII), but since we expect
$n_e\sim n\sim 10$ \cc\/ in this region \citep[see above
and][]{Reach2000}, our $n_e\la n\sim 10^3$ \cc\/ upper limit is not
violated. Unfortunately, previous attempts to model the line emission
from J-type shocks either do not include the infrared fine structure
lines \citep{Dopita1996}, use pre-shock densities too high for this
case \citep[$n_o>10^3$,][]{Hollenbach1989}, or assume that the
pre-shock material is atomic \citep{Hartigan1987} so that these
studies cannot be used to predict the parameters of the 3C~391 in
detail \citep[see][for further discussion of these models with regard
to 3C~391]{Reach2000}.

If an independent estimate for the path length for the ionized gas
responsible for the free-free absorption can be made, we can estimate
the electron density in another way.  \citet{Chen2004} found that the
column density associated with the X-ray absorbing molecular cloud
(i.e. the difference in column between the SE and NW parts of the SNR)
is $6\times 10^{21}$ \cmt\/ (see \S1.1).  Combining this information
with the infrared line post-shock density estimate of $n\sim 10^3$
\cc\/, we find that the path length through the X-ray absorbing gas is
$\sim 2$ pc.  If we assume that this path length is also appropriate
for the ionized gas, and use the average $EM$ derived above of $(0.6 -
2.5)\times 10^3$ cm$^{-6}$ pc, we find that $n_{e,avg}\sim 20-40$
\cc\/. However, there could be a significant contribution to the
molecular cloud column density (determined from the differential x-ray
absorption) by un-shocked gas of lower density than assumed here.
Since it is not possible to quantify this effect (high resolution
\HI\/ absorption measurements would be useful but do not exist), it is
not clear how significant this estimate is. Another independent
assessment of the thickness of the ionized gas layer can be made from
the transverse size scale over which there are significant changes in
the 74~MHz free-free absorption. From Figures 3 and 4, this sizescale
is $1\arcmin-2\arcmin$ which corresponds to 2.2 to 4.4 pc at a
distance of 7.5 kpc. If we make the usual assumption that the depth is
approximately equal to the transverse sizescale, the electron density
is $n_{e,avg}\sim 10-35$ \cc\/ in good agreement with that determined
from the X-ray column density path length.

Electron densities of $10-40$ \cc\/ are consistent with the suggestion
of \citet{Reach2000} that $n_e\sim n/100$ \cc\/, if the post-shock
density is $\sim 10^{3-3.5}$ \cc\/. This range of electron densities
is somewhat higher than typically assumed in other studies of
free-free absorption derived from integrated flux measurements
\citep[e.g. $1-10$ \cc\/][]{Kassim1989b,Dulk1975}.  However, we also 
assume an electron temperature that is somewhat lower: $1,000-3,000$ K
compared to the $3,000-8000$ K typically assumed, so that the implied
pressure is similar. The lower temperature estimated here also seems
reasonable in light of the fact that the $3,000-8000$ K estimate from
\citet{Kassim1989b} is appropriate for the extended envelopes of
\HII\/ regions, i.e. photoionization, compared to the predominately
collisional ionization suggested here for 3C~391.

\subsection{Nature of the 3C~391 74~MHz Absorption}

The nature of the medium that causes free-free thermal absorption
toward the large number of SNRs observed to have low frequency
spectral turnovers remains uncertain \citep[c.f.][]{Kassim1989b}.  For
example, two obvious sources of Galactic ionized gas: the warm ionized
medium (WIM) and \HII\/ regions do not provide the appropriate
physical conditions to account for most of the SNRs with thermal
absorption.  The WIM has electron densities that are too low by a
factor of $\sim 10$ to 100 and its filling factor is too high
\citep[$\sim 20\%$;][]{Kulkarni1987,Reynolds1991}.  This last is
important because \citet{Kassim1989b} showed that SNR free-free
absorption is not correlated with distance, as would be necessary for
a ubiquitous distributed component like the WIM.  Normal \HII\/
regions have electron densities and temperatures that are too high to
account for the moderate low frequency optical depths that are
observed ($\tau_{74}\sim 1$) toward most SNRs with thermal absorption,
although there are a few cases with very high optical depths were this
is a likely scenario (W30 for example).

Two viable alternatives for intermediate temperature and density
ionized thermal gas have been suggested.  \citet{Anantharamaiah1985,
Anantharamaiah1986} has suggested that the extended envelopes of
normal \HII\/ regions (EHEs) might well provide the requisite
temperatures and densities. This suggestion was prompted by the
statistical correlation of high and low frequency RRL velocities,
indicating that they may arise from the same kinematic regions.
However, it is unclear whether such envelopes extend far enough into
the ISM to account for a large fraction of the SNRs showing low
frequency absorption.  Alternatively, \citet{Heiles1996} suggest that
Galactic ``worms'' may cause absorption toward some sources. Ionized
``worms'' occur when a later generation of massive OB stars ionizes a
preexisting superbubble. Indeed, the absorption observed toward the
SNR W49B by \citet{Lacey2001} may be due to such a ``worm''
\citep{Brogan2004}.  However, the physical properties of ``worms'' 
are not currently well determined so it is not clear whether they 
can account for a large number of SNR thermal absorption cases.

{\em The free-free absorption observed toward 3C~391 provides the
first evidence for another origin: ionized gas produced when a SNR
blast wave encounters a nearby molecular cloud with sufficient speed
to dissociate and ionize the gas.} This is an exciting prospect since
it implies that the large fields of view afforded by low radio
frequency observations may be used to efficiently search for cases of
SNR/molecular cloud interactions.  Thus far only a handful of such
cases are known from traditional blind \HI\/ and CO surveys \citep[for
example W44, IC443, W28, and HB21,][and references
therein]{Reach2002}, despite the fact that it seems plausible that
many SNRs should still be located in or near their nascent molecular
clouds. OH (1720 MHz) masers, found toward $\sim 10\%$ of SNRs, have
proven to be powerful signposts of SNR/molecular cloud interactions
\citep{Green1997}. A number of new interaction cases have been
confirmed by follow-up infrared and mm-wavelength observations of SNRs
with OH (1720 MHz) masers, including 3C~391 \citep[also see for
example][]{Lazendic2004,Reynoso2000}. However, these masers require a
very limited set of physical conditions to be pumped \citep[i.e. slow
C-type shocks with $T\sim 100$ K and $n\sim 10^4$
\cc\/][]{Wardle1999}, so that while molecular gas must be present were
they are found, non-detection is inconclusive. Thus, low frequency
free-free thermal absorption may provide an important complementary
tracer of fast ionizing SNR/molecular cloud shocks. Unfortunately,
since typical \HI\/ clouds in the cold neutral medium (CNM) only have
densities of $\sim 0.5$ \cc\/ \citep[see for example][]{Heiles2003},
it is unlikely that SNR/CNM \HI\/ clouds interactions would produce
observable free-free absorption.

One obvious benefit to be gained by finding more SNR/molecular cloud
interactions, is that the velocity of the cloud can then be used to
derive a kinematic distance for the remnant. In this regard, radio
recombination line detections of the free-free ionized gas would also
yield velocity information. SNRs distances are notoriously difficult
to obtain although it is a key parameter in most SNR diagnostics like
physical size, age, and energetics. Given the low number of SNRs for
which accurate distance estimates exist, even a few new free-free
absorption driven SNR/molecular cloud discoveries would represent a
significant gain. For currently known examples of SNR/molecular cloud
interaction, free-free absorption also represents an important means
of studying the properties of the shock, since only J-type shocks can
create the requisite ionized gas. However, it is also notable that if
the free-free absorption observed toward 3C~391, is common toward SNRs
with associated molecular clouds, the observation of intrinsic
spectral index variation due to shock acceleration will be very
difficult to detect at low radio frequencies.

\section{CONCLUSIONS}

For the first time we have been able to spatially resolve (at
$70\arcsec$ resolution) the morphology of the low frequency free-free
absorption toward 3C~391, previously inferred from integrated flux
measurements.  We find that the average 74~MHz optical depth is 1.1
and the spectral index of the SNR is $-0.49$.  There is impressive
agreement between the regions of strongest 74~MHz free-free absorption
and ionized fine structure atomic lines between 12-18 \mum\/. This
coincidence suggests that the source of ionized thermal gas
responsible for the low frequency absorption arises from the
SNR/molecular cloud shock boundary. The appropriate range of
temperatures and densities to excite the fine structure lines can be
easily achieved in the post-shock gas of a J-type shock with a speed
of $\sim 100$ \kms\/ \citep{Reach2002}. Using the range of
temperatures appropriate for the infrared ionized fine structure lines
of $1,000-3,000$ K \citep{Reach2002} we find that the average emission
measure of the free-free absorbing gas is $EM =(0.6 - 2.5)\times 10^3$
cm$^{-6}$~pc.  From a number of different lines of evidence including
the infrared fine structure emission and the observed differential
X-ray absorption across 3C~391, we find that the electron density
likely lies between $10-10^3$ \cc\/, with a lower range between 
$10-40$ \cc\/ favored. 

A study by \citet{Kassim1989b} suggested that $\sim 2/3$ of all SNRs
in the Galaxy suffer from some degree of low frequency free-free
absorption. It has been proposed that such absorption is due to a {\em
diffuse} ($n_e\sim 1-10$ \cc) ionized component of the ISM - possibly
the extended envelopes of \HII\/ regions. This new work shows for the
first time that such absorption can arise from ionized gas in the
post-shock region behind a SNR/molecular cloud shock interface. This
result is in direct contrast to the result of \citet{Lacey2001} for
the SNR W49B where the low frequency absorption was shown to be linked
to an unrelated kinematic component along the line of
sight. Additionally, the electron temperatures ($\sim~1,000-3,000$ K)
associated with the thermal ionized gas surrounding 3C391 is lower
than the canonical values ($3,000 - 8,000$ K) assumed by previous
interpretations of thermal ISM absorption \citep[see for
example][]{Kassim1989b}. 

Our result implies that thermal absorption effects towards
SNR/molecular cloud complexes may be significantly stronger than
previously suspected, if J-type shocks are common.  This is only the
second case after W49B for which {\em extrinsic} free-free thermal
absorption toward an SNR has been spatially resolved (free-free
absorption from thermal gas interior to a remnant has also been
observed toward CasA and the Crab).  Future spatially resolved VLA
74~MHz studies toward a larger sample of Galactic SNRs from our
ongoing VLA 74~MHz inner Galactic plane survey ($l =-10\arcdeg$ to
$50\arcdeg$; Brogan et al. 2004, in prep.) will be a powerful tool in
the search for elusive SNR/molecular cloud interactions. They will
also allow us to study the prevalence of the other possible sources of
free-free absorption described in \S4.4. The next generation of low
frequency instruments like the Long Wavelength Array (LWA) and the Low
Frequency Array LOFAR, with their superior resolution and sensitivity
will likely shed even greater light on this phenomenon.

\acknowledgements

We would like to thank S. P. Reynolds for giving us permission to
utilize his archival VLA 330 and 1465 MHz data. We would also like to
thank W. Reach for supplying us with his CO and IR data in electronic
form.  Likewise for P. Slane for supplying us with the X-ray data. 
We would also like to thank the anonomous referee for helping us to 
improve the clarity of the text. 
Basic research in radio astronomy at the NRL is supported by the
Office of Naval Research. KKD acknowledges support during this
research as an NSF Astronomy and Astrophysics Postdoctoral Fellow
under award AST-0103879 and currently as a National Research Council
Postdoctoral Fellow.

\begin{figure}[h!]
\epsscale{1.0}
\plotone{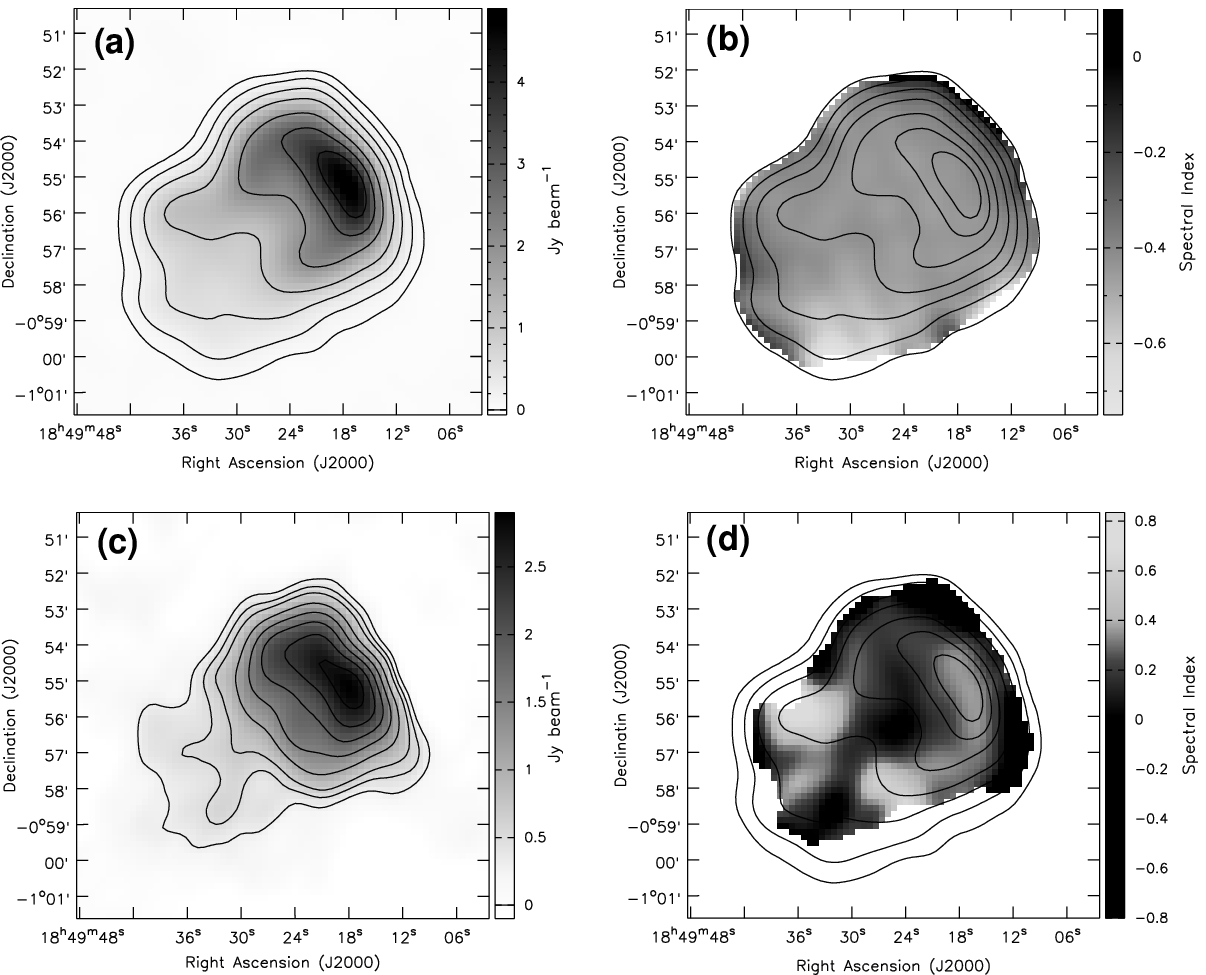}
\caption[]{(a) VLA 330~MHz continuum image of 3C~391 with $70\arcsec$
resolution and contours at 0.06, 0.2, 0.5, 1, 2, 3, \& 4
\jb\/. The rms noise of this image is 0.015 \jb\/.  (b) Spectral index
map of 3C~391 between 330/1465 MHz. The 330~MHz continuum contours from
(a) are superposed. The spectral index is quite uniform with an
average value of $-0.45$. (c) VLA 74~MHz continuum image of 3C~391 
with contours at 0.3, 0.5, 0.7, 1.1, 1.5, 1.9, 2.3, and 2.7 \jb\/ 
(the rms noise is 0.075 \jb\/).
(d) Spectral index map of 3C~391 between
74/330~MHz. The 330~MHz continuum contours from (a) are
superposed. Note that the spectral indices are inverted (positive)
over much of the map (the grey scale is also inverted compared to 
(b)). }
\end{figure}

\begin{figure}[h!]
\epsscale{0.5}
\plotone{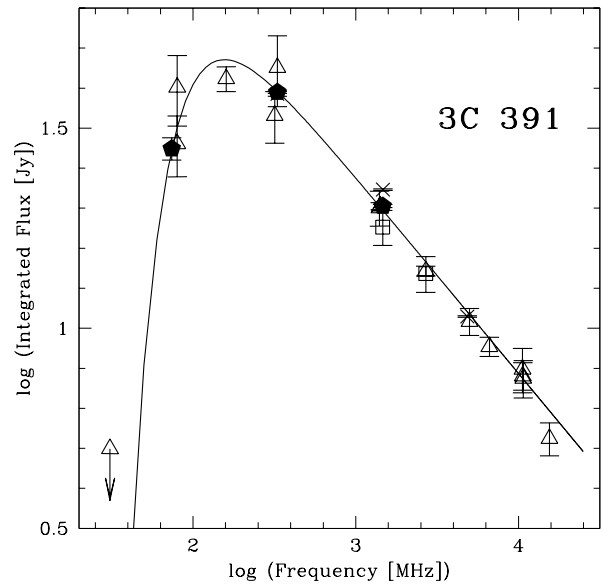}

\caption[]{Radio continuum spectrum for SNR 3C~391. The solid line is
the fit to the data using Equation 1 excluding the 30.9 MHz upper
limit. The fitted spectral index is $-0.49\pm 0.01$.
The filled hexagon symbols are from the current work, the
$\bigtriangleup$ symbols are from \citet{Kassim1989a, Kassim1992}, the
$\square$ symbols are from \citet{Reich2001} and \citet{Reich1997},
while the $\times$ symbols are from \citet{Moffett1994}. }

\end{figure}

\begin{figure}[h!]
\epsscale{1}
\plotone{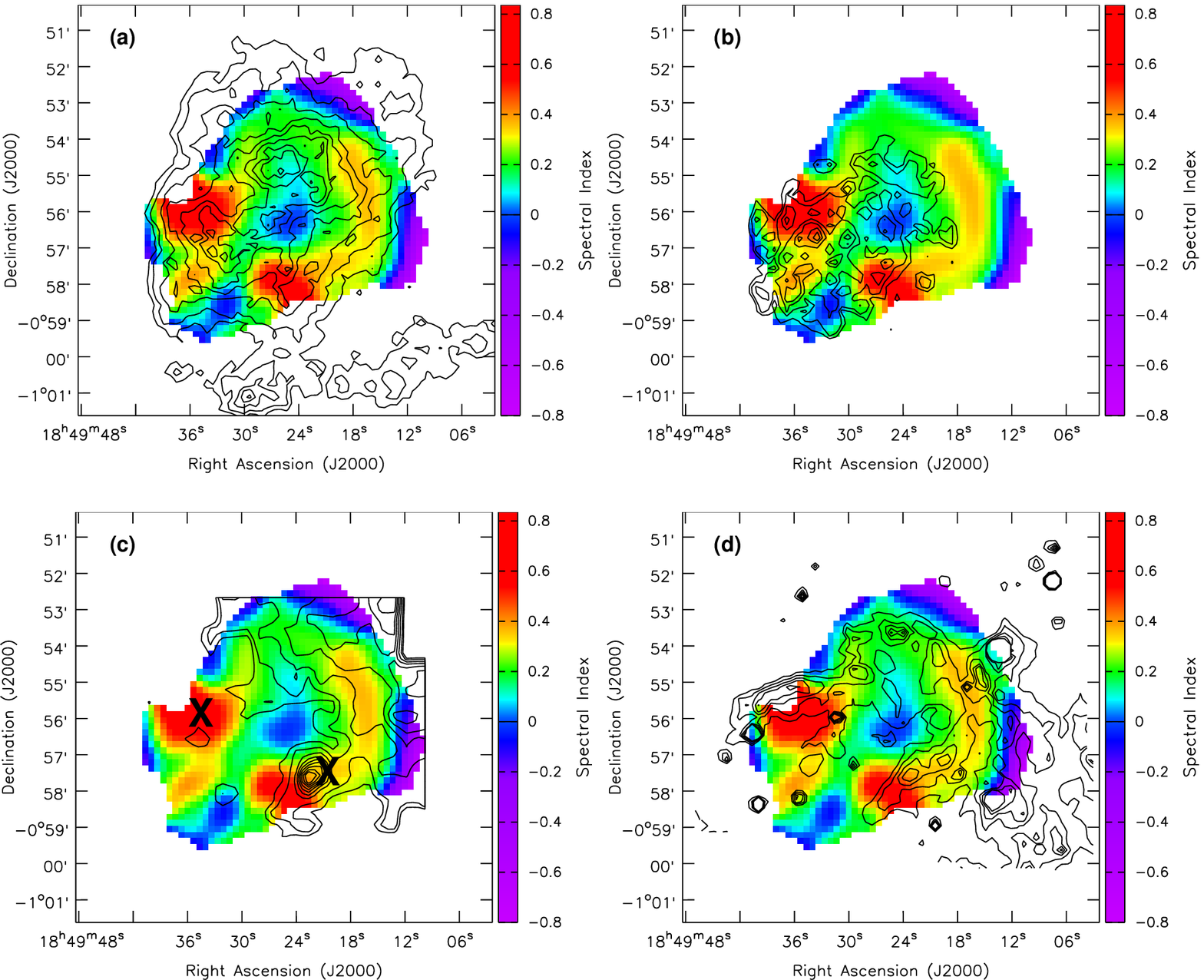}

\caption[]{Color version of the 74/330~MHz spectral index map
presented in Figure 1d superposed with (a) Hard X-ray contours from
ASCA in the 2.6-10 keV range \citet{Chen2001}; (b) Soft X-ray contours
from ASCA in the 0.5-2.6 keV range \citep{Chen2001}; (c) CO (2-1)
integrated emission contours from 91 to 110 \kms\/ tracing the molecular cloud
interacting with 3C 391 \citep{Reach1999}; and (d) IR emission
contours from ISOCAM at 12-18 \mum\/ \citep{Reach2002}. The $\times$
symbols on (c) indicate the positions of OH (1720 MHz) masers
\citep{Frail1996}.  There is impressive agreement between the
regions of strongest 74~MHz absorption and the IR emission. }

\end{figure}

\begin{figure}[h!]
\epsscale{0.5}
\plotone{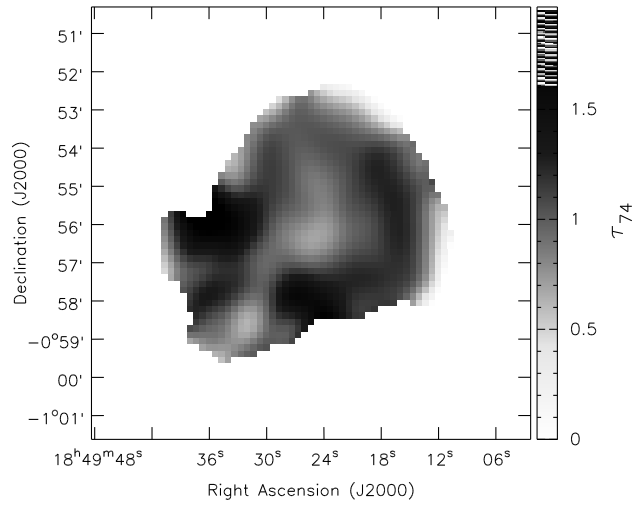}

\caption[]{Optical depth toward 3C~391 at 74~MHz as a function of
position. The $\tau_{74}$ optical depth was calculated using Equation
2, assuming a constant spectral index of $-0.49$. Using this map
along with an estimate of the electron temperature, the 74~MHz
emission measure ($EM$) can be calculated from Equation 3
($EM=0.051~\tau_{74}~T_e^{1.35}$ cm$^{-6}$ pc).}

\end{figure}

\begin{deluxetable}{lccr}
\small
\tablewidth{25pc}
\tablecaption{VLA Observational Parameters\label{tab1}}
\tablecolumns{4}
\tablehead{
\colhead{Date}  & \colhead{Config.} &\colhead{Bandwidth} 
& \colhead{ Time$^a$} 
\\ &   &\colhead{(MHz)} & \colhead{(Hours)}}
\startdata
\cutinhead{74~MHz parameters}

2000 Dec 31 & A &  1.5 & 1.33\\
2001 Jan 13 & A & 1.5 & 1.33 \\
2001 Mar 01 &  B & 1.5 & 1.7 \\
2001 Aug 28 & C & 1.5 & 1.0 \\

\cutinhead{330~MHz parameters$^b$}
1994 Oct 23 & C  & $3.1\times$ 2 & 2.5 \\
\cutinhead{1465 MHz parameters$^c$}
1991 Apr 12 & D & $6.25\times$ 2 & 1.25 \\

\enddata
\tablenotetext{a} {Approximate time on source}
\tablenotetext{b} {Archival VLA data from project AR311}
\tablenotetext{c} {Archival VLA data from project AR232}
\end{deluxetable}

\end{document}